\documentclass[journal,twoside,web]{ieeecolor}
\usepackage{generic}
\usepackage{cite}
\usepackage{amsmath,amssymb,amsfonts}
\usepackage{algorithmic}
\usepackage{graphicx}
\usepackage{algorithm,algorithmic}
\usepackage{hyperref}
\hypersetup{hidelinks=true}
\usepackage[caption=false,font=footnotesize]{subfig}
\usepackage{textcomp}
\usepackage{booktabs}
\usepackage{tabularx}
\usepackage{makecell}
\usepackage{multirow}
\usepackage{array}   
\usepackage[normalem]{ulem}
\def\BibTeX{{\rm B\kern-.05em{\sc i\kern-.025em b}\kern-.08em
    T\kern-.1667em\lower.7ex\hbox{E}\kern-.125emX}}
\begin{document}
\title{TinyUSFM: Towards Compact and Efficient Ultrasound Foundation Models}
\author{Chen Ma, Jing Jiao, Shuyu Liang, Junhu Fu, Qin Wang, Zeju Li, Yuanyuan Wang, \IEEEmembership{Senior Member, IEEE}, and Yi Guo, \IEEEmembership{Member, IEEE}
\thanks{This work was supported by National Key R\&D Program of China (2024YFF0507300, 2024YFF0507303), National Natural Science Foundation of China (Grant No. 62531004), and Shanghai Municipal Education Commission (Grant No. 24KNZNA09) (Corresponding authors: Yuanyuan Wang; Yi Guo).}
\thanks{Chen Ma, Jing Jiao, Shuyu Liang, Junhu Fu, Qin Wang, Zeju Li, Yuanyuan Wang, and Yi Guo are with the College of Biomedical Engineering, Fudan University, Shanghai 200433, China (email: \parbox[t]{0.48\textwidth}{cma24@m.fudan.edu.cn; jiaojing@fudan.edu.cn; syliang22@m.fudan.\\edu.cn; jhfu21@m.fudan.edu.cn; wangqin23@m.fudan.edu.cn; zejuli@fudan.edu.cn; yywang@fudan.edu.cn; guoyi@fudan.edu.cn).}}}

\maketitle

\begin{abstract}
Foundation models for medical imaging demonstrate superior generalization capabilities across diverse anatomical structures and clinical applications. Their outstanding performance relies on substantial computational resources, limiting deployment in resource-constrained clinical environments. This paper presents TinyUSFM, the first lightweight ultrasound foundation model that maintains superior organ versatility and task adaptability of our large-scale Ultrasound Foundation Model (USFM) through knowledge distillation with strategically curated small datasets, delivering significant computational efficiency without sacrificing performance. Considering the limited capacity and representation ability of lightweight models, we propose a feature-gradient driven coreset selection strategy to curate high-quality compact training data, avoiding training degradation from low-quality redundant images. To preserve the essential spatial and frequency domain characteristics during knowledge transfer, we develop domain-separated masked image modeling assisted consistency-driven dynamic distillation. This novel framework adaptively transfers knowledge from large foundation models by leveraging teacher model consistency across different domain masks, specifically tailored for ultrasound interpretation. For evaluation, we establish the UniUS-Bench, the largest publicly available ultrasound benchmark comprising 8 classification and 10 segmentation datasets across 15 organs. Using only 200K images in distillation, TinyUSFM matches USFM's performance with just 6.36\% of parameters and 6.40\% of GFLOPs. TinyUSFM significantly outperforms the vanilla model by 9.45\% in classification and 7.72\% in segmentation, surpassing all state-of-the-art lightweight models, and achieving 84.91\% average classification accuracy and 85.78\% average segmentation Dice score across diverse medical devices and centers. This work successfully bridges the gap between high-performance foundation models and practical clinical deployment, winning the first place in MICCAI2025 IUGC Challenge.
\end{abstract}

\begin{IEEEkeywords}
Ultrasound image analysis, Lightweight foundation model, Task adaptability
\end{IEEEkeywords}

\section{Introduction}
\label{sec:introduction}
\IEEEPARstart{U}{ltrasound} imaging is one of the most widely used diagnostic modalities owing to its safety, affordability, and real-time capability~\cite{intro1}. As it plays a central role in routine clinical practice, developing reliable automated ultrasound analysis systems is crucial for improving diagnostic efficiency and reducing operator dependence. Deep learning-based approaches have achieved remarkable progress in various ultrasound tasks such as organ segmentation and disease classification~\cite{intro2}. However, achieving robust and generalizable automated ultrasound analysis still remains highly challenging, as image appearance is strongly influenced by operator experience, probe placement, and imaging protocol. Slight variations in probe angle or position can lead to significant structural differences even in images of the same organ. Furthermore, acquisitions from different ultrasound machines or vendors often introduce substantial discrepancies in contrast, resolution, and noise patterns. These factors result in weak generalizability of automatic analysis systems across patients, devices, and clinical settings.

Recently, medical foundation models have emerged as powerful tools for learning generalizable representations from large-scale data, demonstrating promising results across diverse clinical applications~\cite{intro3}. In particular, Ultrasound Foundation Model (USFM)~\cite{usfm} demonstrated that large-scale pretraining with spatial-frequency dual masking can produce a generalizable encoder for multiple tasks and organs. The outstanding performance of large foundation models relies on architectures with hundreds of millions of parameters, demanding significant computational and memory resources. This poses a major barrier to their practical deployment in real-world clinical settings, particularly in resource-constrained scenarios. Such practical gap motivates the development of lightweight yet high-performance ultrasound foundation models, which aim to preserve the representational capacity of large-scale foundation models while improving their feasibility for widespread clinical use~\cite{intro4}.

Existing studies have explored model compression through architectural simplification (e.g., MobileViT~\cite{mobilevit}, EfficientNet~\cite{efficientnet}) and knowledge distillation~\cite{B1} from large teachers to lightweight students. The key objective is to preserve representational power and task performance while drastically reducing parameters and computation. Building lightweight ultrasound foundation models that are both efficient and effective still faces two critical challenges. First, large-scale ultrasound datasets exhibit strong heterogeneity and contain a considerable proportion of low-quality or redundant samples~\cite{intro5}. Directly training lightweight models on such uncurated data can lead to underfitting or spurious feature learning~\cite{intro6}. Second, maintaining strong performance under strict parameter constraints requires effective knowledge transfer, yet most existing distillation methods are designed for natural images and single tasks~\cite{intro7}, thus fail to capture the complex spectral and spatial characteristics of ultrasound~\cite{intro8} or build a foundation model.

To address these challenges, we propose TinyUSFM, a lightweight ultrasound foundation model that retains the versatility and adaptability of our large-scale USFM while requiring far fewer computational resources. The model employs a feature-gradient driven coreset selection strategy to curate diverse and informative samples for efficient training, and incorporates a consistency-driven dynamic distillation assisted by domain-separated masked image modeling(MIM), which leverages teacher consistency across spatial and frequency masks to guide reliable knowledge transfer while mitigating negative transfer from ambiguous samples. The main contributions of this work are summarized as follows:

\begin{itemize}
\item[$\bullet$] We present TinyUSFM, the first lightweight ultrasound foundation model specifically designed to maintain broad organ versatility and task adaptability of large-scale models while drastically reducing computational and memory costs. It provides a practical solution for deploying foundation-level intelligence on resource-constrained clinical platforms.
\item[$\bullet$] We design a unified framework combining feature–gradient driven coreset selection with consistency-driven knowledge distillation, augmented by domain-separated masked image modeling. The former ensures diverse, high-quality, and training-beneficial data curation from large ultrasound datasets, while the latter enables reliable and domain-consistent knowledge transfer that preserves both spatial and frequency representations.
\item[$\bullet$] To facilitate comprehensive and standardized evaluation, we construct UniUS-Bench, the largest publicly available ultrasound benchmark covering 15 organs with 8 classification and 10 segmentation tasks from diverse medical centers and imaging devices. Extensive experiments demonstrate that TinyUSFM outperforms all state-of-the-art lightweight models on several challenging tasks, enabling reliable and efficient clinical deployment.
\end{itemize}

\section{Related Work}
\subsection{Medical Foundation Models}
Foundation models have revolutionized medical AI by establishing generalist systems capable of handling diverse tasks~\cite{intro3}. Their success largely stems from large-scale pretraining on millions of heterogeneous images using masked image modeling or contrastive learning to capture rich semantic representations transferable across organs, modalities, and tasks~\cite{A2}. Several domain-specific foundation models have been developed following this paradigm. For example, MIS-FM leverages masked autoencoding to learn universal representations from computed tomography volumes~\cite{A3}, RETFound employs vision transformers pretrained on millions of retinal images for fundus analysis~\cite{A4}, Endo-FM integrates temporal attention to capture spatial–temporal correlations in endoscopy videos~\cite{A5}, and MedSAM extends the general SAM architecture to medical image segmentation~\cite{A6}. Most notably, Jiao et al. proposed USFM, the first universal ultrasound foundation model trained on the 3M-US database, which contains over two million ultrasound images collected from multiple organs, centers, and devices, employing spatial-frequency dual masking to address ultrasound analysis challenges~\cite{usfm}.

Despite their strong performance, existing medical foundation models are extremely large, demanding substantial computational resources and hindering deployment in resource-constrained clinical environments~\cite{A7}. This practical limitation has motivated research on lightweight foundation models that preserve generalization capacity while remaining feasible for real-world clinical use~\cite{intro4}.

\subsection{Knowledge Distillation}
A widely adopted approach to constructing lightweight models is knowledge distillation (KD), which transfers information from high-capacity teacher models to lightweight student networks~\cite{B1}. Initially applied to image classification using soft logits~\cite{B2}, KD has been extended to dense prediction tasks such as semantic segmentation by leveraging structured pixel-level information~\cite{B3}. In medical imaging, KD has proven useful for compressing models while retaining accuracy. Qin et al. introduced an efficient segmentation architecture by distilling from well-trained networks to lightweight ones~\cite{B4}. Subsequent works explored deep mutual distillation for semi-supervised segmentation to improve performance on ambiguous regions~\cite{B5}, multi-domain mutual distillation for enhanced representation across datasets~\cite{B6}, cross-layer graph flow distillation~\cite{B7}, and generalizable knowledge distillation~\cite{B8}. While prior KD studies in medical imaging have achieved success in compressing single-task networks, extending them to foundation models is non-trivial. Foundation models must transfer broad, multi-organ and multi-task knowledge rather than task-specific features, and ultrasound data add further difficulty with strong speckle noise, shadowing, and device variability. These factors make generic distillation strategies ineffective and highlight the need for ultrasound-oriented, foundation-level approaches.

\subsection{Coreset Selection}
Foundation models are typically pretrained on massive and heterogeneous datasets that inevitably contain a considerable proportion of low-quality or redundant samples. While large teacher models possess sufficient capacity to effectively absorb useful knowledge from such uncurated data, lightweight student models with limited capacity are more susceptible to underfitting or learning spurious features. Coreset selection offers a promising solution by extracting representative and informative subsets, thereby improving both training efficiency and model robustness~\cite{C1}. 

\begin{figure*}[!t]
    \centering
    \includegraphics[width=0.95\linewidth]{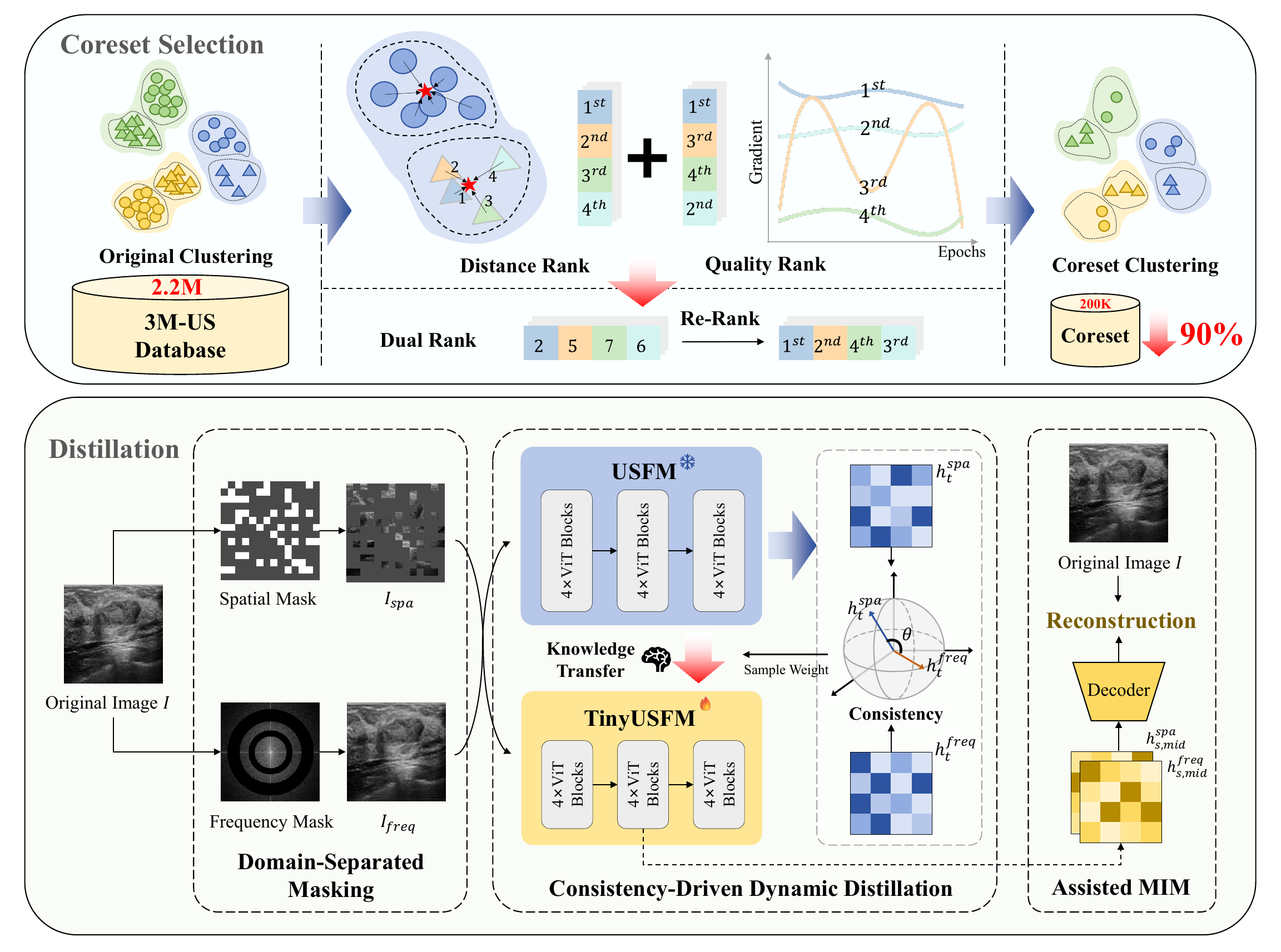}
    \caption{Overview of proposed TinyUSFM.}
    \label{fig1}
\end{figure*}

Traditional approaches adopt various selection criteria, such as prototype distance~\cite{C2}, gradient norm~\cite{C3}, influence function scores~\cite{C4}, and uncertainty metrics~\cite{C5}, often combined with clustering algorithms (e.g., K-means) to mitigate redundancy~\cite{C6}. However, coreset selection in medical imaging poses unique challenges. Medical datasets commonly exhibit severe class imbalance and long-tailed distributions~\cite{C7}, and medical foundation models are expected to capture diverse data characteristics across multiple organs and imaging devices. Existing methods, designed primarily for general-purpose vision tasks, rarely account for these domain-specific factors. Moreover, coreset selection strategies specifically tailored for foundation model construction and knowledge distillation in medical imaging remain largely unexplored.

\section{Methodology}
\subsection{Method Overview}
Fig.~\ref{fig1} illustrates the overview of our TinyUSFM framework. It transfers knowledge from the high-capacity USFM(86.5M) to lightweight TinyUSFM(5.5M) through two key components. First, a feature-gradient driven coreset selection extracts diverse and high-quality samples from USFM’s pretraining 3M-US database. Second, a domain-separated MIM assisted consistency driven dynamic distillation performs deep-layer knowledge transfer combined with mid-level spatial and frequency reconstruction. The resulting TinyUSFM serves as a versatile lightweight feature encoder, enabling both efficient downstream adaptation and effective feature extraction while preserving high-quality ultrasound representations.

\subsection{Feature-Gradient Driven Coreset Selection}
Lightweight models are particularly sensitive to noisy or redundant samples, which are prevalent in large-scale heterogeneous ultrasound datasets. To ensure effective and efficient knowledge distillation under limited model capacity, we propose a feature-gradient driven coreset selection strategy that curates the most training-beneficial samples while maintaining balanced, diverse, and high-quality representation. 

Our strategy jointly leverages feature diversity and training contribution, two complementary factors that capture data representativeness and learnability. This dual-drive principle combines: (1) feature-driven clustering, which hierarchically organizes the teacher model (USFM) feature space to preserve inter-organ and inter-device diversity; and (2) gradient-driven quality assessment, which quantifies each sample’s learning value based on gradient stability and magnitude during teacher pretraining.

To capture multi-scale balance and diversity without overrepresenting dominant organs or devices, we embed all images $I_i$ in the 3M-US database using the teacher encoder $\phi_T(\cdot)$ to obtain feature representations $z_i=\phi_T(I_i)$. A two-level hierarchical K-means structure is then applied, and both levels follow the same optimization objective function below:
\begin{equation}
\begin{aligned}
\underset{C,\,A}{\min} \;
& \sum_{i=1}^{N} \sum_{j=1}^{K} a_{i,j} \| z_i - c_j \|^2, \\[3pt]
\text{s.t. } \;
& a_{i,j} \in \{0,1\}, \;
\sum_{j=1}^{K} a_{i,j} = 1.
\end{aligned}
\end{equation}
where $N$ denotes the number of images and $K$ denotes the number of clusters. $C=\{c_j\}_{j=1}^K$ are cluster centroids and $A=[a_{i,j}]$ is the assignment matrix. The first level partitions data by major anatomical structures, while the second refines intra-organ variations across imaging devices.

While clustering ensures diversity, it alone does not guarantee learning utility. We therefore introduce a gradient-based criterion to estimate each sample’s training value based on teacher’s optimization. Using gradients collected from the first ten epochs in our previous USFM pretraining, we compute mean $\mu_i$ and variance $\sigma_i$ of gradients for each sample, and derive two complementary metrics: Stability score,
\begin{equation}
    s_i^{stb}=1-\frac{\sigma_i}{\mu_i},
\end{equation}
measuring the temporal consistency, and magnitude score,
\begin{equation}
    s_i^{mag}=\frac{\mu_i-min_j\{\mu_j\}}{max_j\{\mu_j\}-min_j\{\mu_j\}},
\end{equation}
representing the relative contribution to model updates. Their weighted sum 
\begin{equation}
    s_i=\alpha s_i^{stb}+(1-\alpha)s_i^{mag},
\end{equation}
where $\alpha$ is empirically set to 0.5 to balance stability and magnitude, evaluates both reliability and contribution. Samples with large and stable gradients reflect regions where the teacher learns generalizable representations, aligning data curation with the teacher’s knowledge landscape.

Finally, we integrate feature diversity and gradient quality through a dual-ranking strategy. Within each fine-grained cluster, we compute the Euclidean distance from each sample to the cluster centroid. Each sample obtains two ranks, $r_i^Q$ and $r_i^D$, according to its descending quality score and ascending Euclidean distance, respectively. The final selection score is calculated as 
\begin{equation}
    r_i=\beta r_i^Q+(1-\beta)r_i^D,
\end{equation}
where $\beta$ is set to 0.5 to equally weight quality and diversity. The top-k samples from each subcluster constitute the final coreset, ensuring both representational coverage and learning efficiency.

\subsection{Domain-Separated MIM assisted Consistency-Driven Dynamic Distillation}

To effectively distill knowledge from USFM to TinyUSFM, we propose a novel consistency-driven dynamic distillation framework assisted by domain-separated MIM. The core of our approach is to assess the teacher model's reliability based on its predictive consistency under spatial and frequency perturbations, thereby adaptively adjusting the distillation weight. Complementing this process, the domain-separated MIM explicitly preserves the mid-level spatial and spectral representations that are crucial for ultrasound image analysis.

\subsubsection{Domain-Separated Masking Strategy}
We first introduce a domain-separated masking strategy that independently perturbs ultrasound images in spatial and frequency domains to generate complementary views for both reliability assessment and representation learning. This design is motivated by the distinct roles of each domain: the spatial domain primarily captures structural and textural information, while the frequency domain encodes spectral and speckle patterns inherent to ultrasound imaging. Such a separation enables more effective evaluation of teacher model consistency and facilitates targeted, domain-specific reconstruction. 

For each input ultrasound image $I$, the strategy generates two distinct masked views.

Spatial masking replaces randomly selected patches with the image’s mean intensity:
\begin{equation}
M_s(I) =
\begin{cases}
mean(I), & (x, y) \in \textit{masked patches}\\[4pt]
I(x, y), & \textit{otherwise.}
\end{cases},
\end{equation}
where $(x,y)$ are the coordinates in the spatial domain, and $mean(I)$ is calculated across the entire image to ensure the continuity of the grayscale distribution. This operation encourages the model to infer missing anatomical structures from contextual cues.

Frequency masking adopts a band-stop filter applied to the magnitude spectrum while preserving the phase component:
\begin{equation}
M_f(I) = \mathcal{F}^{-1}\!\big( BS(|\mathcal{F}(I)|; f_1, f_2) \cdot e^{j\phi(u,v)} \big),
\end{equation}
where $\mathcal{F}$ and $\mathcal{F}^{-1}$ denote the Fourier and inverse Fourier transforms, $BS(\cdot)$ defines the band-stop range between $f_1$ and $f_2$, and $\phi(u,v)$ is the preserved phase. This perturbation suppresses specific frequency bands, compelling the model to recover missing spectral information.

\subsubsection{Consistency-Driven Dynamic Distillation}
\begin{figure*}[htbp]
    \centering
    \includegraphics[width=0.95\linewidth]{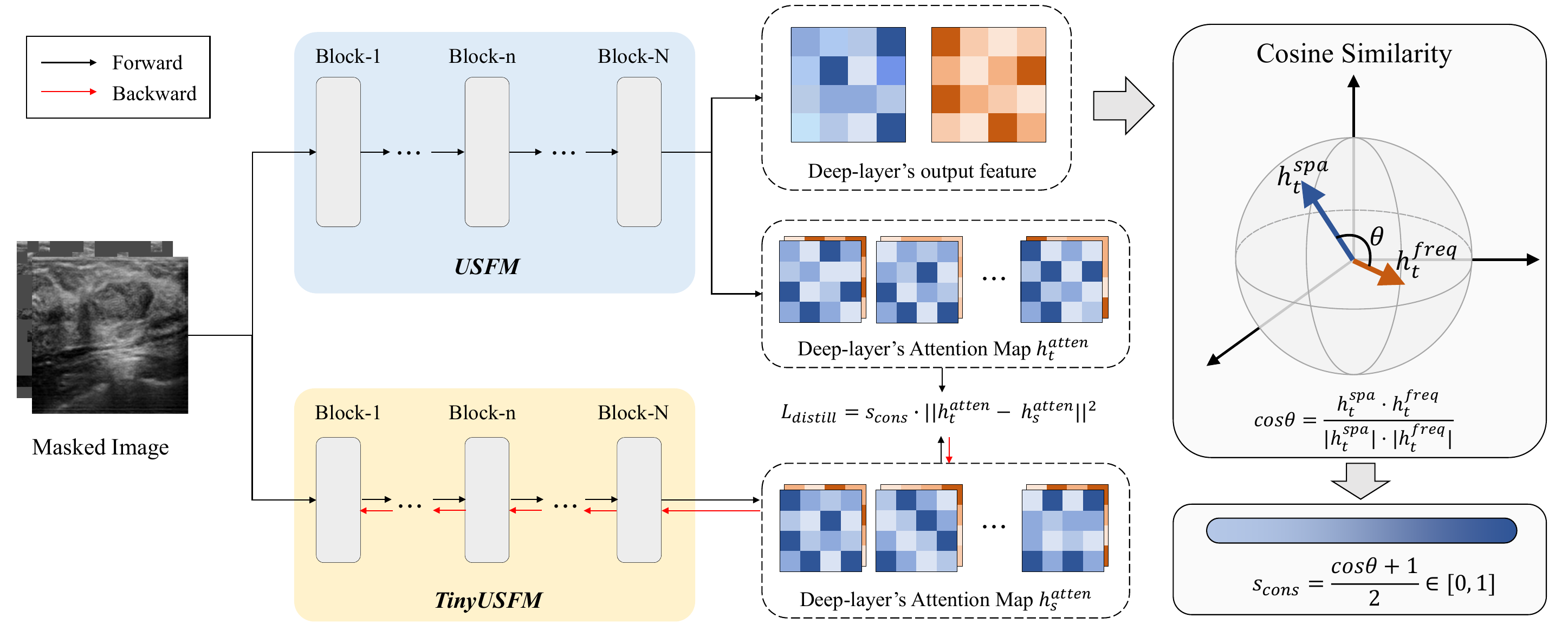}
    \caption{Details of Consistency-Driven Dynamic Distillation.}
    \label{fig2}
\end{figure*}

Building upon the domain-separated masking strategy, we develop a consistency-driven dynamic distillation framework that adaptively regulates knowledge transfer, as illustrated in Fig.~\ref{fig2}. The key idea is to evaluate the reliability of the teacher model by measuring how consistently it represents an image when different spatial or frequency masks are applied. If the deep-layer features extracted from these two masked views remain highly similar, it indicates that the teacher has captured stable and meaningful ultrasound representations. Such samples are regarded as reliable and assigned higher weights during distillation, whereas samples with inconsistent responses are down-weighted to avoid misleading supervision. This mechanism enables the student to focus on trustworthy knowledge, thereby achieving more effective and robust representation learning.

Given the domain-separated masked inputs $I_{spa}$ and $I_{freq}$ obtained by the masking strategy, the teacher model $F_t(\cdot)$ extracts deep-layer representations as: 
\begin{equation}
h_t^{spa}=F_t(I_{spa}), \quad h_t^{freq}=F_t(I_{freq}).
\end{equation}

The teacher consistency score is computed based on cosine similarity between two representations: 
\begin{equation}
    s_{cons}=\frac{cos(h_t^{spa},h_t^{freq})+1}{2},
\end{equation}
where $cos(\cdot)$ denotes the cosine similarity function. A higher $s_{cons}$ indicates that the teacher encodes similar semantics under different perturbations, implying a more stable and trustworthy knowledge source.  

The consistency score dynamically weights the distillation loss applied to the deep layers. The weighting is performed via attention head feature alignment, which ensures that reliable samples exert a stronger supervisory influence, while ambiguous ones are appropriately down-weighted. To enable direct head-wise feature alignment and preserve representational granularity, TinyUSFM uses a modified ViT-Tiny architecture with the same number of attention heads as the teacher USFM. This design allows one-to-one head correspondence between student and teacher for precise feature matching. Formally, the dynamic distillation loss is defined as:
\begin{equation}
\mathcal{L}_{distill} = s_{cons} \cdot 
\frac{1}{H} \sum_{h=1}^{H} 
\text{MSE}\left(h^{atten,h}_{s}, \, h^{atten,h}_{t}\right),
\end{equation}
where $h_s^{atten,h}$ and $h_t^{atten,h}$ represent the features from the h-th attention head for student and teacher respectively.

This mechanism adaptively emphasizes reliable supervision, ensuring an efficient and stable knowledge transfer from the teacher.

\subsubsection{Domain-Separated MIM}
While deep-layer feature alignment effectively transfers high-level semantic knowledge, it risks causing the lightweight student to overlook crucial mid-level representations that capture fine-grained structural and spectral cues essential for ultrasound understanding. To compensate for this, we introduce a domain-separated masked image modeling (MIM) scheme that provides targeted mid-level supervision through self-reconstruction in spatial and frequency domains. By separating two reconstruction pathways, TinyUSFM can focus on learning structural textures and spectral patterns within its limited capacity, achieving stable convergence and more comprehensive knowledge transfer.

The student encoder processes both spatial and frequency masked views and extracts intermediate features $h_{spa}^{mid}$ and $h_{freq}^{mid}$ at the mid-layer for two complementary objectives. A lightweight decoder is then applied to recover the original inputs, which enables the student to learn informative representations through self-supervised reconstruction. Specifically, the reconstructed outputs are obtained as follows:
\begin{equation}
\hat{I}_{spa} = F_{D}(h_{spa}^{mid}), \quad 
\hat{I}_{freq} = F_{D}(h_{freq}^{mid}).
\end{equation}
where $F_D$ denotes a standard MAE-style masked reconstruction decoder. The combined objective for mid-level layer reconstruction is then defined as:
\begin{equation}
\mathcal{L}_{recon} = 
\text{MSE}(\hat{I}_{spa}, I) + 
\lambda\cdot\text{MSE}(\hat{I}_{freq}, I),
\end{equation}
where $\lambda=1.0$ balances spatial vs frequency reconstruction terms.

During training, the student backbone and the decoder $F_D$ are optimized jointly with a single total objective:
\begin{equation}
\mathcal{L}_{total} = \mathcal{L}_{distill} + \eta \mathcal{L}_{recon},
\end{equation}
where $\eta$ is used to balance the reconstruction term against the distillation term and is set to 0.5. The teacher is kept frozen and only provides supervision signals.

This integrated framework enables TinyUSFM to efficiently acquire both spatial and spectral representations while maintaining computational efficiency, allowing the student model to effectively absorb the teacher’s knowledge and achieve robust, high-fidelity representations for ultrasound image analysis.

\begin{table*}[!ht]
\centering
\caption{Summary of the UniUS-Bench, including 8 classification and 10 segmentation datasets across 15 organs.}
\label{tab1}
\renewcommand{\arraystretch}{1.1}
\setlength{\tabcolsep}{3pt}
\begin{tabular}{m{1.9cm} m{1.6cm} m{3.2cm} m{2.2cm} m{1.1cm} m{4.5cm}}
\toprule
Downstream & Category & Dataset & Organs & \#Images & Targets \\
\midrule
\multirow{8}{*}{Classification} 
 & \multirow{4}{*}{Binary} 
   & CUBS\,\cite{CUBS} & Carotid & 1378 & Non-FUP / FUP Event \\
 &  & UF1990\,\cite{UF1990} & Uterine & 1990 & Normal / Fibroid \\
 &  & TN3K\,\cite{TN3K} & Thyroid & 3491 & Benign / Malignant \\
 &  & STMUS\,\cite{STMUS} & Skeletal muscle & 5312 & Normal / Pathological \\[2pt]
 & \multirow{4}{*}{Multi-class}
   & AUL\,\cite{AULliver} & Liver & 735 & Normal / Benign / Malignant \\
 &  & BUSI\,\cite{busi} & Breast & 780 & Normal / Benign / Malignant \\
 &  & MMOTU\,\cite{MMOTU} & Ovarian & 1469 & 8 Subtype tumors \\
 &  & Fetal Planes\,\cite{fetalplanes} & Fetus & 12400 & 6 Fetal planes \\
\midrule
\multirow{11}{*}{Segmentation} 
 & \multirow{7}{*}{Single-target}
   & Luminous\,\cite{luminous} & Multifidus muscle & 341 & Lumbar multifidus muscle \\
 &  & KidneyUS\,\cite{kidneyus} & Kidney & 487 & Kidney capsule \\
 &  & GIST514\,\cite{usd514} & Stomach & 514 & Gastrointestinal stromal tumor \\
 &  & DDTI\,\cite{ddti} & Thyroid & 637 & Thyroid nodule \\
 &  & MMOTU\,\cite{MMOTU} & Ovarian & 1469 & Ovarian tumor \\
 &  & BUSBRA\,\cite{busbra} & Breast & 1875 & Breast tumor \\[2pt]
 &  & Ultrasound Nerve Seg.\,\cite{ultrasound-nerve-segmentation} & Neck Nerve & 5735 & Brachial Plexus \\[2pt]
 & \multirow{3}{*}{Multi-target}
   & LUSS\,\cite{luss} & Lung & 564 & Rib / Pleural line / A-line / B-line / B-line confluence \\
 &  & FH-PS-AoP\,\cite{fhpsaop} & Pelvis & 4000 & Pubic symphysis / Fetal head \\
 &  & CAMUS\,\cite{camus} & Cardiac & 19232 & LV cavity / LV myocardium / LA \\
\bottomrule
\end{tabular}
\end{table*}

\section{Experiments}
\subsection{Datasets}
Our study utilizes two purpose-specific datasets. For pretraining, we apply our feature-gradient driven coreset selection strategy to curate a 200K subset from the large-scale 3M-US database~\cite{usfm} to enable efficient knowledge distillation. 

For evaluation, we introduce UniUS-Bench, the largest publicly available ultrasound benchmark designed for standardized and comprehensive assessment of ultrasound foundation models. Unlike previous datasets that are often limited to a single organ or task, UniUS-Bench integrates diverse clinical scenarios from different medical centers and imaging devices into a unified evaluation suite. It encompasses 8 classification and 10 segmentation datasets, covering 15 organs with a total of 60,940 ultrasound images and 62,409 annotations. Importantly, UniUS-Bench is built entirely from existing public datasets without including any private data, ensuring openness, transparency, and reproducibility, thereby making it the most extensive and clinically relevant benchmark to date. 

The benchmark includes diagnostic classification tasks covering both binary and multi-class organ-level diagnosis (e.g., liver, breast, fetal planes), as well as segmentation tasks spanning single-target and multi-target structures at both organ-level (e.g., kidney, fetal head) and tumor-level (e.g., breast tumor, thyroid nodule, liver lesion). This comprehensive design enables systematic evaluation of model generalization across diverse anatomical structures, imaging devices, and clinical tasks.

Detailed data statistics, including sample size, task type, and organ coverage, are summarized in Table~\ref{tab1}. To enable fair comparison, we follow the official training, validation and testing splits whenever available. For those datasets without predefined splits, we adopt a stratified random partition with a ratio of 7:1:2 for training, validation, and testing, respectively.

\subsection{Compared methods}
To comprehensively evaluate TinyUSFM, we compare it with three categories of representative methods, including conventional models, lightweight models, and foundation models.

For the classification task, we consider the following methods: (1) conventional CNN/Transformer models: ResNet~\cite{resnet} and ViT-B~\cite{vitb}, (2) lightweight models specifically designed for efficiency: MobileViT~\cite{mobilevit}, EfficientNet~\cite{efficientnet}, and ViT-Tiny~\cite{deit}, and (3) foundation models: URFM~\cite{urfm} and our teacher model USFM~\cite{usfm}. Model performance is reported in terms of classification accuracy.

For the segmentation task, we evaluate TinyUSFM against the following methods: (1) conventional segmentation models: VM-UNet~\cite{vmunet} and RWKV-UNet~\cite{rwkvunet}; (2) lightweight segmentation models: SegFormer~\cite{segformer}, SeaFormer~\cite{seaformer}, and ViT-Tiny with FPN decoders, specifically designed for efficiency; and (3) foundation models: URFM and the teacher model USFM. Model performance is reported in terms of Dice score.

All conventional and lightweight models are initialized with ImageNet pretrained weights, while USFM and URFM are initialized with their ultrasound pretrained weights.

\subsection{Implementation details}
All input images were resized to 224×224. During pretraining, the spatial masking ratio was set to 0.75. For frequency masking, we followed the USFM configuration by defining seven band-pass filters spanning from low to high frequencies and randomly selecting two filters to generate the mask. A 10×10 area at the center of the frequency spectrum, containing crucial low-frequency information, was always preserved. The frequency masking ratio was set to 0.4. 

For downstream tasks, training employed a warmup poly learning rate scheduler with an initial learning rate of 1e-4 for 400 epochs. Data augmentation strategies were tailored to each task. For classification, we employed random flipping, rotation, and scaling. For segmentation, we additionally employed contrast and gamma adjustments to enhance model robustness.

All results are reported as mean ± std. For classification, std is computed over five runs with different random seeds, whereas for segmentation it is calculated across test images using per-image metrics.

All experiments were conducted on an AMD EPYC 7763 CPU and NVIDIA A100 GPU. All models were developed using PyTorch. All compared methods were implemented strictly following their official repositories or the settings reported in their original papers to ensure fair comparison.

\section{Results}

We present a comprehensive evaluation of TinyUSFM, assessing its pre-training efficacy and downstream adaptability. The evaluation first analyzes how the distillation framework enhances representation quality and efficiency. It then benchmarks TinyUSFM against conventional, lightweight, and foundational models across diverse tasks. Together, these experiments provide a holistic evaluation of how TinyUSFM achieves efficient and reliable knowledge transfer while maintaining lightweight deployment.

\subsection{Effectiveness of TinyUSFM pretraining}

To comprehensively validate the effectiveness of TinyUSFM pretraining, we conduct systematic analyses from multiple perspectives. Specifically, we first perform ablation studies on the overall framework to assess the contribution of each proposed component, followed by targeted studies on coreset selection efficiency, domain-separated reconstruction superiority and consistency-driven dynamic distillation benefit. We further employ UMAP~\cite{umap} for representation visualization to examine TinyUSFM’s discriminability and generalization across organs and tasks. Finally, we compare the model size and computational cost to highlight the compactness and efficiency of our TinyUSFM. 

\begin{table}[htbp]
\centering
\caption{Overall framework ablation study. Results are reported as classification accuracy (\%) and segmentation Dice score (\%) on UniUS-Bench. Best results are in \textbf{bold}.}
\label{tab:ablation}
\begin{tabular}{
>{\centering\arraybackslash}m{1cm}
>{\centering\arraybackslash}m{1cm}
>{\centering\arraybackslash}m{1cm} c c}
\toprule
\multicolumn{3}{c}{Components} & \multirow{2}{*}{Classification} & \multirow{2}{*}{Segmentation} \\
Coreset & MIM & Distillation & & \\
\midrule
$\times$ & $\times$ & $\times$ & 75.46 & 78.06 \\
$\times$ & $\times$ & $\checkmark$ & 81.93 & 82.03 \\
$\times$ & $\checkmark$ & $\checkmark$ & 83.54 & 84.01 \\
$\checkmark$ & $\times$ & $\checkmark$ & 83.50 & 84.59 \\
$\checkmark$ & $\checkmark$ & $\checkmark$ & \textbf{84.91} & \textbf{85.78} \\
\bottomrule
\label{tab2}
\end{tabular}
\end{table}

Table~\ref{tab2} presents comprehensive ablation results evaluating the contribution of each component: feature-gradient driven coreset selection, domain-separated MIM, and consistency-driven dynamic distillation. The baseline configuration without any components achieves 75.46\% average classification accuracy and 78.06\% average segmentation Dice score on UniUS-Bench. Adding consistency-driven dynamic distillation alone provides substantial improvements to 81.93\% and 82.03\% respectively, demonstrating the effectiveness of dynamic distillation weighting and the strong representational capability of USFM. With the addition of domain-separated MIM as an auxiliary task, distillation achieves enhanced performance of 83.54\% and 84.01\%, highlighting the importance of mid-level reconstruction learning. Incorporating feature-gradient coreset selection with distillation yields 83.50\% accuracy and 84.59\% Dice score, demonstrating that selecting high quality and non-redundant data benefits model distillation by reducing noisy supervision. The complete framework integrating all three components achieves optimal performance with 84.91\% classification accuracy and 85.78\% segmentation Dice score, validating the synergistic effect of combining these three components.

\subsubsection{Efficiency of Coreset Selection}

\begin{figure}[!t]
    \centering
    \subfloat[Ablation on subset size.]{%
        \includegraphics[width=0.24\textwidth]{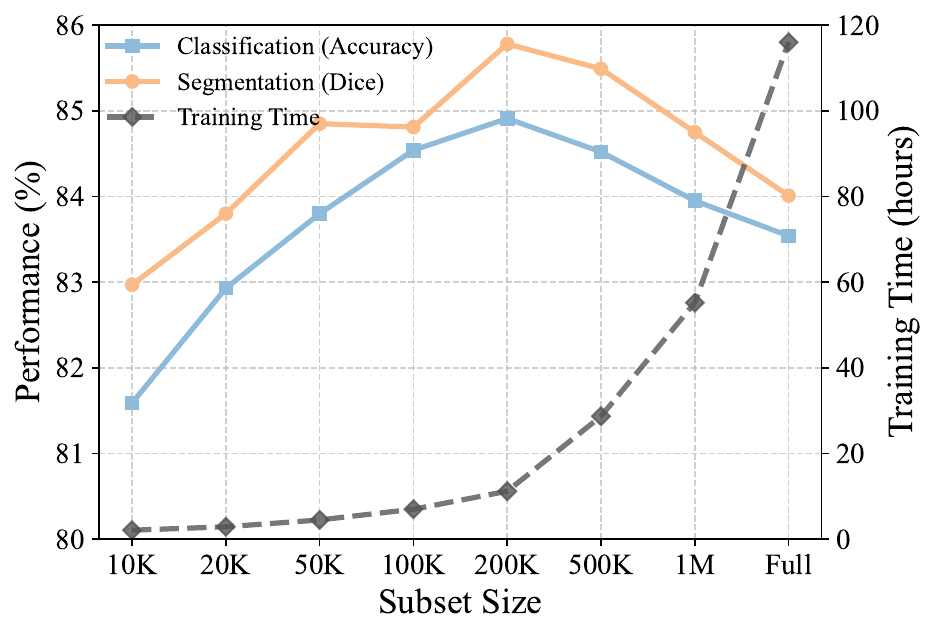}%
        \label{fig3a}}
    \hfill
    \subfloat[Comparison at 200K subset.]{%
        \includegraphics[width=0.24\textwidth]{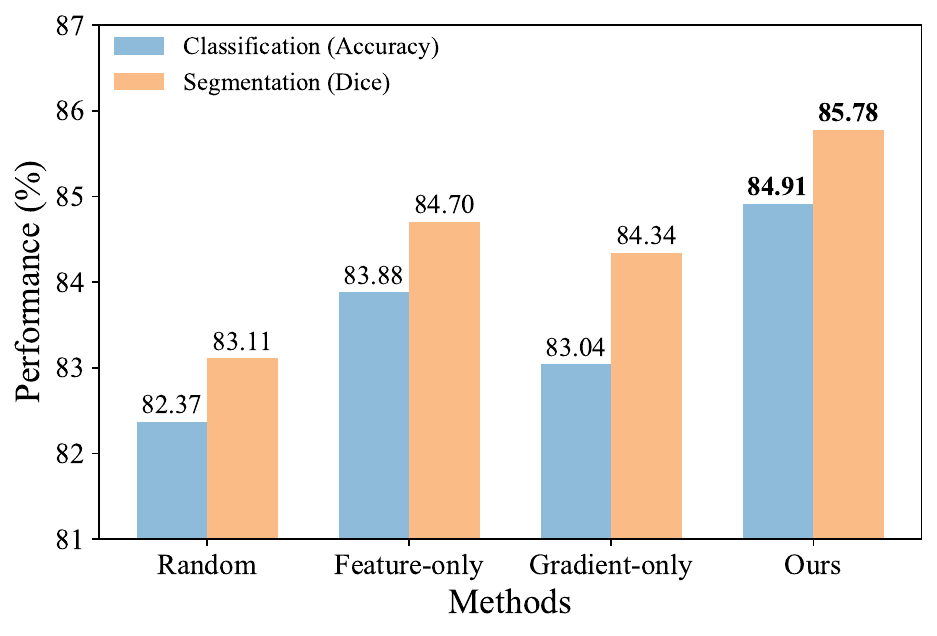}%
        \label{fig3b}}
    \caption{Ablation study on coreset selection strategy.}
    \label{fig3}
\end{figure}

Fig.~\ref{fig3} illustrates the ablation study conducted to evaluate our coreset selection strategy. We investigate the impact of subset size by constructing training sets ranging from 10K to the full 3M-US dataset containing over 2.1 million images. This experiment investigates the relationship between dataset size and its effects on the performance of lightweight models and the efficiency of distillation. In addition, after determining 200K as the optimal subset size, we further compare our feature–gradient driven selection against several alternative strategies, including random sampling, feature-only clustering, and gradient-only quality filtering. This design allows us to assess both the optimal subset size and the relative advantage of combining diversity preservation with quality-based ranking.

The results reveal two key observations. First, performance consistently improves as the subset size grows from 10K to 200K, with average classification accuracy rising from 81.59\% to 84.91\% and average segmentation Dice score from 82.97\% to 85.78\%. Notably, the optimal 200K subset, using merely 10\% of the full dataset, outperforms training on the complete 3M-US database by 1.37\% in classification and 1.77\% in segmentation, while requiring less than 10\% of the training time on a single NVIDIA A100 GPU. This finding demonstrates that curated high-quality data can be more effective and efficient than large redundant datasets for lightweight models. Beyond 200K, both accuracy and Dice score begin to decline (83.54\% and 84.01\% at full dataset) and the training time increases substantially, suggesting that additional low-quality samples not only compromise representation quality but also introduce unnecessary computational overhead. 

Second, our integrated approach significantly outperforms all baselines at the 200K subset: achieving +2.54\% in classification accuracy and +2.67\% in segmentation Dice over random sampling, +1.03\% and +1.08\% over feature-only clustering, and +1.87\% and +1.44\% over gradient-only filtering, respectively. These findings underscore that effective lightweight model training relies not merely on data reduction, but on carefully curated data that balances diversity with quality, thereby validating the design of our coreset selection method.

\subsubsection{Superiority of Domain-Separated Reconstruction}

To further validate the superiority of the proposed domain-separated MIM, we conduct comprehensive analyses of the best reconstruction layer and the reconstruction domain configuration.

\begin{figure}[!t]
    \centering
    \subfloat[Performance variation across different reconstruction layers.]{%
        \includegraphics[width=0.24\textwidth]{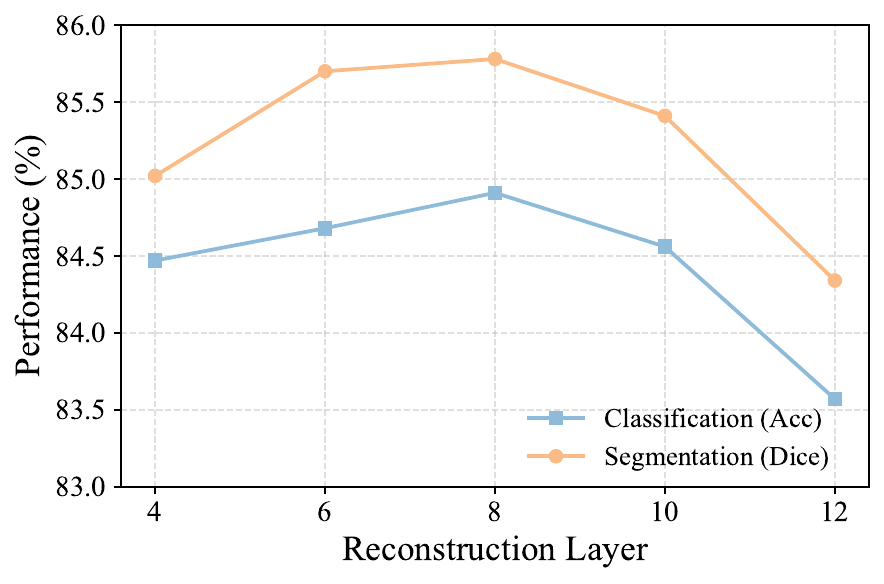}%
        \label{fig4a}}
    \hfill
    \subfloat[Comparison of reconstruction domain configurations.]{%
        \includegraphics[width=0.24\textwidth]{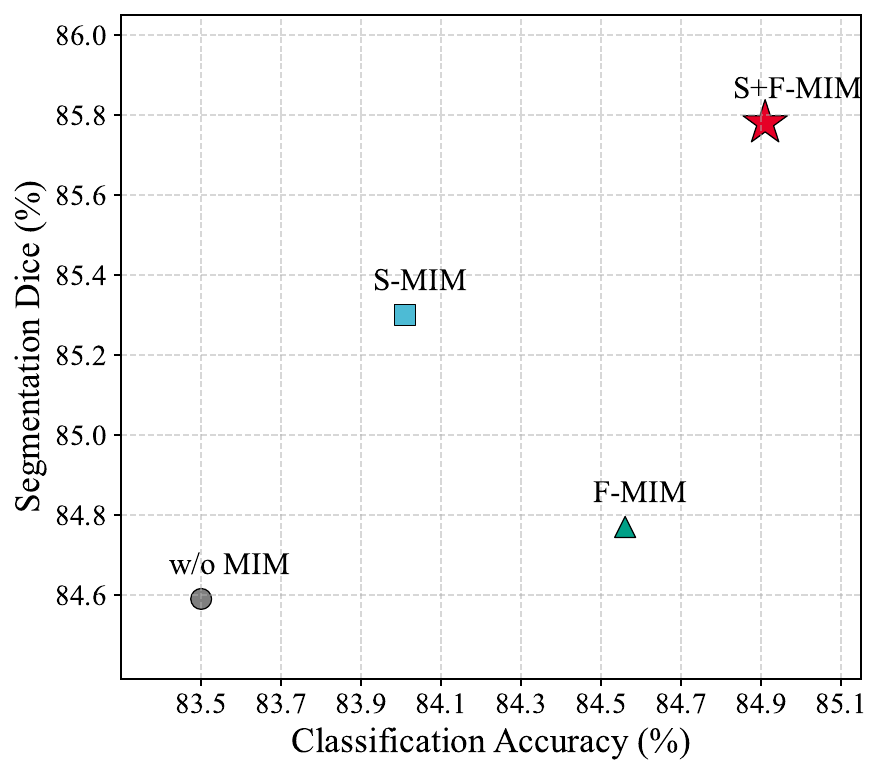}%
        \label{fig4b}}
    \caption{Ablation studies on domain-separated reconstruction.}
    \label{fig4}
\end{figure}

\begin{figure*}[!t]
    \centering
    \subfloat[Distribution of 3M-US Database in TinyUSFM feature space.]{%
        \includegraphics[width=0.48\textwidth]{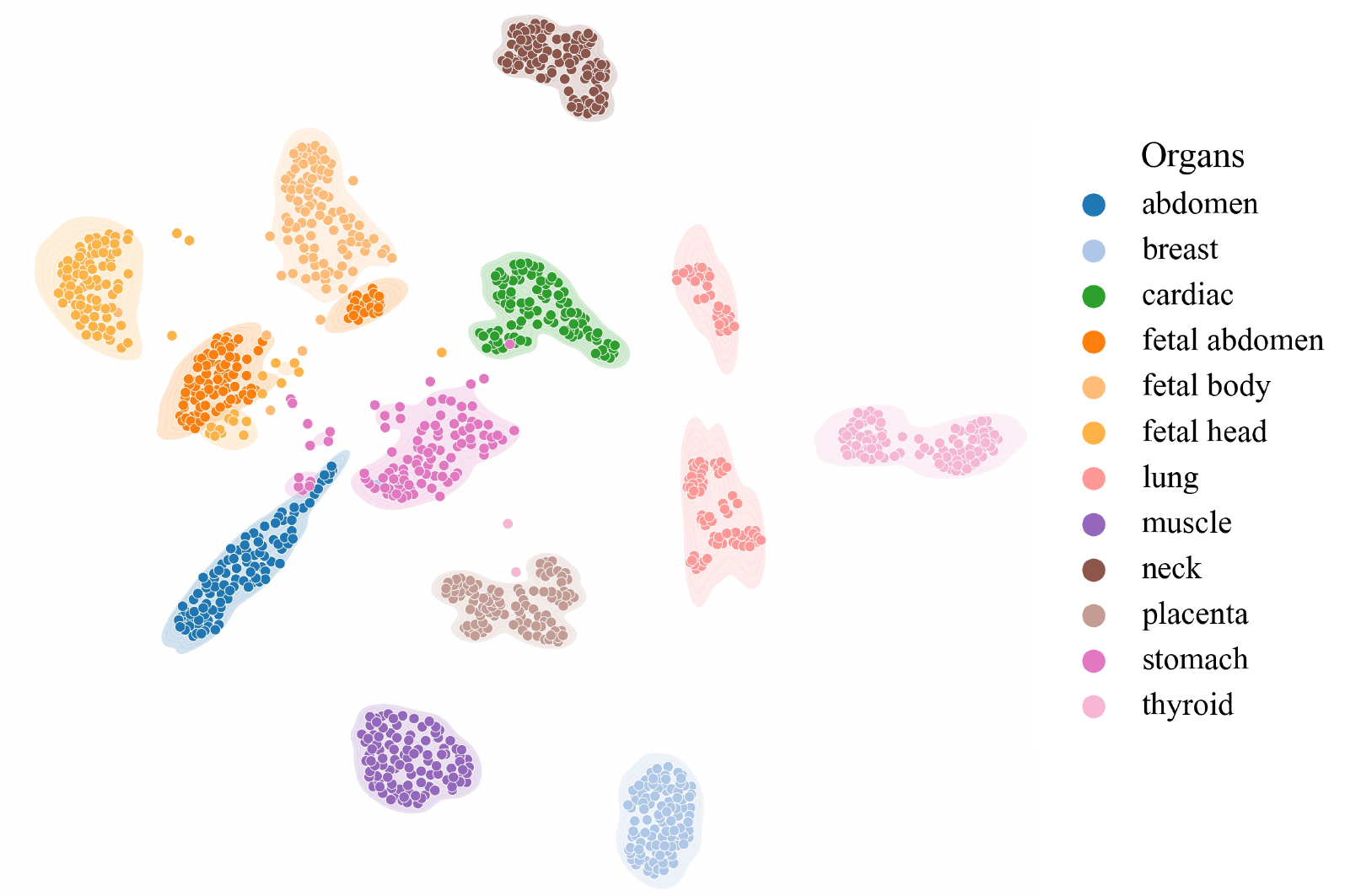}%
        \label{fig5a}}
    \hfill
    \subfloat[Distribution of UniUS-Bench in TinyUSFM feature space.]{%
        \includegraphics[width=0.48\textwidth]{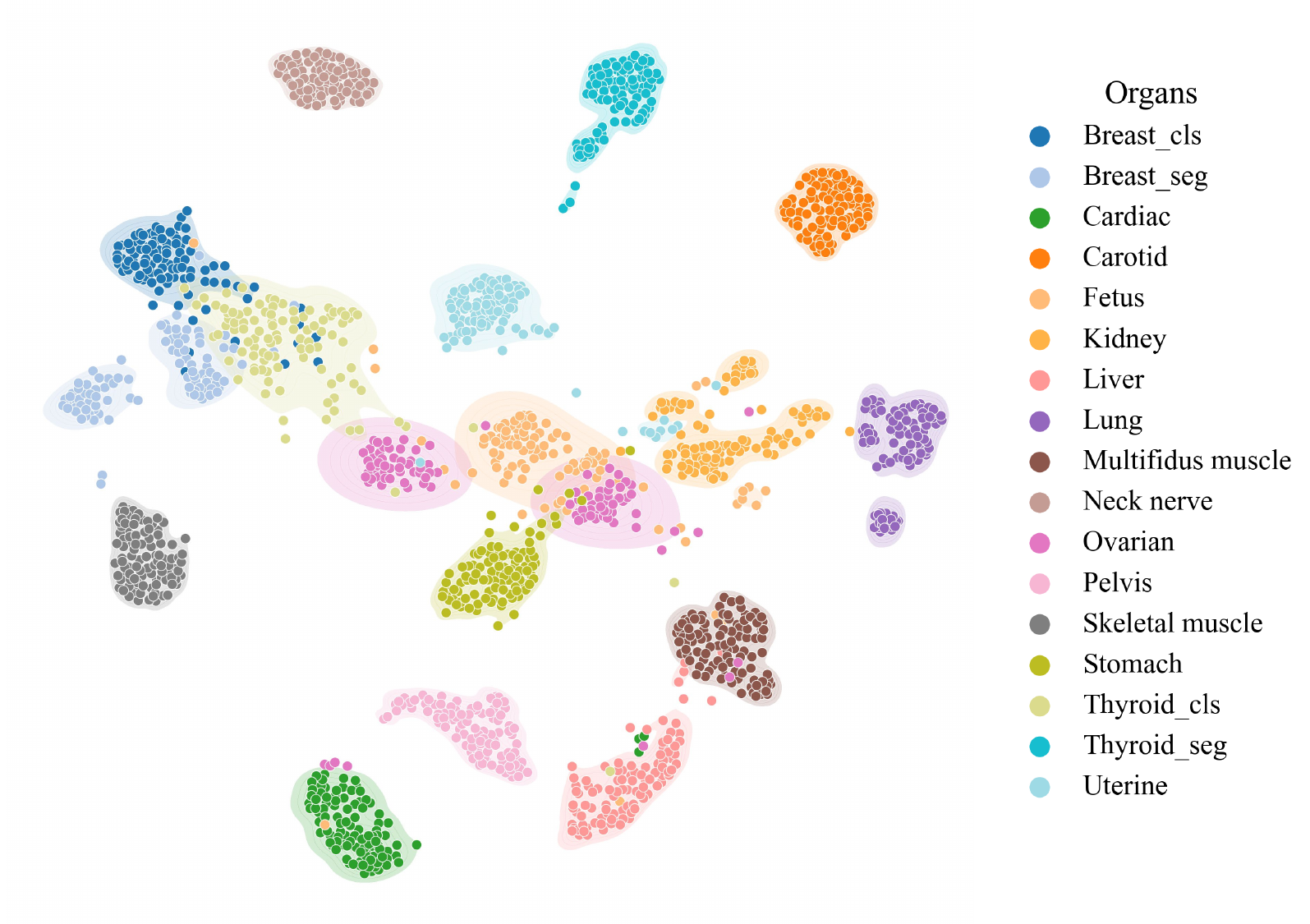}%
        \label{fig5b}}
    \caption{UMAP visualization of TinyUSFM feature representations.}
    \label{fig5}
\end{figure*}

Fig.~\ref{fig4a} investigates the optimal layer for applying domain-separated MIM reconstruction in our distillation framework. We evaluate reconstruction at different layers while keeping other components fixed. Both classification and segmentation tasks show consistent trends, with performance gradually improving from shallow layers to mid-level layers, peaking at the 8th layer with 84.91\% average classification accuracy and 85.78\% average segmentation Dice score. Notably, performance significantly degrades when applying reconstruction at the deep-layer for both tasks, dropping to 83.57\% in classification and 84.34\% in segmentation. This suggests that performing reconstruction and deep-layer feature alignment simultaneously at the same layer creates conflicting objectives that impair learning effectiveness. These findings validate our design strategy of placing reconstruction at the mid-layer and feature alignment at the deep-layer, which achieves optimal performance while avoiding optimization conflicts between different learning objectives.

In addition, we examine the effect of different reconstruction domains, including spatial-only (S-MIM), frequency-only (F-MIM), and spatial–frequency combined (S+F-MIM) configurations, alongside the baseline without MIM. As illustrated in Fig.~\ref{fig4b}, the baseline without MIM achieves 83.50\% classification accuracy and 84.59\% segmentation Dice score. Incorporating spatial reconstruction (S-MIM) raises performance by +0.51\% and +0.71\%, while frequency reconstruction (F-MIM) further improves it by +1.06\% and +0.18\%, confirming that both domains contribute complementary information beneficial for ultrasound representation learning. The domain-separated MIM achieves the highest performance, significantly outperforming both single-domain variants. This superiority arises because spatial reconstruction enhances structural and texture perception, whereas frequency reconstruction captures spectral and noise-related characteristics unique to ultrasound imaging.

Together, these findings demonstrate that performing mid-level reconstruction with domain-separated MIM offers the most effective and stable configuration. This design allows the lightweight TinyUSFM to capture comprehensive spatial–spectral representations, thereby improving the deep-layer knowledge transfer quality and overall model robustness.

\subsubsection{Benefit of Consistency-Driven Dynamic Distillation}
We further compare our Consistency-Driven Dynamic Distillation with representative distillation baselines under the same ViT-Tiny student and USFM teacher, as shown in Table~\ref{tab:kd_baselines}. We consider three baselines: feature-based distillation~\cite{B2} that matches the final-layer features via MSE with a lightweight MLP projection head and discards it after pretraining; relation-based distillation following TinyMIM~\cite{tinymim} that aligns token relations derived from $QK^\mathrm{T}$ and $VV^\mathrm{T}$ using KL divergence; and attention-based distillation without the consistency constraint by setting $s_{cons}=1$. For a controlled comparison, we keep the coreset selection and the MAE auxiliary learning unchanged and only replace the distillation term.

Our method achieves the best overall performance on both task types. It reaches 84.91\% average accuracy on classification and 85.78\% average Dice on segmentation, improving over the strongest baseline by 1.32\% and 1.82\%, respectively. The gains are consistent across diverse organs and tasks, indicating that the proposed distillation mechanism provides robust benefits for learning compact ultrasound representations. Moreover, as all methods share the same strong teacher and training pipeline, the observed performance gains cannot be attributed solely to the presence of teacher supervision. Instead, they arise from how the distillation signal is constructed and weighted.

\begin{table}[t]
  \centering
  \caption{Comparison with distillation baselines. Results are reported as classification accuracy (\%) and segmentation Dice score (\%) on UniUS-Bench. Best results are in \textbf{bold}.}
  \label{tab:kd_baselines}
  \begin{tabular}{lcccc}
    \toprule
    Task &
    \makecell[c]{Feature-\\based} &
    \makecell[c]{Relation-\\based} &
    \makecell[c]{Attention-\\based} &
    Ours \\
    \midrule
    Classification & 81.07 & 83.49 & 83.59 & \textbf{84.91} \\
    Segmentation   & 81.60 & 83.64 & 83.96 & \textbf{85.78} \\
    \bottomrule
  \end{tabular}
\end{table}

\subsubsection{Discriminability and Generalizability of Learned Feature Representations}

To qualitatively analyze the feature representations learned during TinyUSFM pretraining, we randomly sample 100 images per organ from both the curated 3M-US pretraining subset and the UniUS-Bench. The embeddings of these samples are visualized using UMAP, enabling inspection of organ-level distribution and similarity patterns in the learned latent space. By projecting high-dimensional features to two dimensions, we assess how well the pretraining captures inter-organ separability and whether the representations generalize across downstream classification and segmentation tasks.

Fig.~\ref{fig5} demonstrates that TinyUSFM successfully learns organ-specific representations during pretraining. In the 3M-US feature space (Fig.~\ref{fig5a}), different organs form distinct, well-separated clusters with clear boundaries, indicating that the model captures meaningful anatomical differences across ultrasound imaging of various body parts, thereby exhibiting strong discriminability of organ-level representations. Each organ type occupies a distinct region in the feature space, demonstrating effective discrimination between different anatomical structures.

\newcommand{\second}[1]{\textbf{#1}}
\newcommand{\best}[1]{\uline{\textbf{#1}}}
\begin{table*}[!htbp]
\centering
\small
\caption{Classification performance comparison on UniUS-Bench. Results are reported as accuracy (\%). Best results are in \uline{\textbf{bold-underlined}}, second best are in \textbf{bold}. Latency is reported as per-image latency at batch size = 64.}
\begin{tabularx}{\textwidth}{l *{10}{>{\centering\arraybackslash}X}}
\toprule
Types & \multicolumn{2}{c}{Conventional Model} & \multicolumn{3}{c}{Lightweight Model} & \multicolumn{3}{c}{Foundation Model} \\
 \cmidrule(lr){2-3} \cmidrule(lr){4-6} \cmidrule(lr){7-9}
Models & ResNet50 & ViT-B & EfficientNet & MobileViT & ViT-T & URFM & USFM & TinyUSFM \\
\midrule
\#Params & 23.5M & 86.5M & 7.7M & 4.4M & 5.5M & 86.5M & 86.5M & 5.5M \\
GFLOPs & 8.21 & 33.72 & 1.34 & 2.86 & 2.16 & 33.72 & 33.72 & 2.16 \\
Latency & 0.31ms & 2.17ms & 0.31ms & 0.36ms & 0.39ms & 2.17ms & 2.17ms & 0.39ms   \\
\midrule
Carotid  & 77.54$\pm$1.54 & 81.16$\pm$1.24 & 78.99$\pm$0.89 & 81.88$\pm$0.96 & 81.88$\pm$0.96 & 83.52$\pm$0.58 & \best{86.23$\pm$0.40} & \second{84.32$\pm$0.67} \\
Uterine  & 91.67$\pm$0.28 & 89.39$\pm$0.63 & 93.43$\pm$0.27 & 95.20$\pm$0.28 & 90.66$\pm$0.28 & 94.95$\pm$0.22 & \best{96.72$\pm$0.22} & \second{96.21$\pm$0.19} \\
Thyroid  & 64.27$\pm$1.31 & 65.91$\pm$1.23 & 67.05$\pm$1.33 & 64.11$\pm$1.09 & 59.05$\pm$2.51 & 70.96$\pm$0.84 & \best{72.59$\pm$0.67} & \second{71.45$\pm$0.71} \\
Muscle   & 87.18$\pm$0.41 & 86.42$\pm$0.59 & 89.36$\pm$0.45 & 89.17$\pm$0.47 & 88.32$\pm$0.49 & 91.74$\pm$0.30 & \best{92.59$\pm$0.29} & \second{91.93$\pm$0.28} \\
Liver    & 74.32$\pm$0.89 & 70.27$\pm$1.16 & 74.32$\pm$0.93 & 72.30$\pm$0.95 & 68.92$\pm$1.20 & 76.35$\pm$0.79 & \best{78.38$\pm$0.75} & \second{77.70$\pm$0.83} \\
Breast   & 83.33$\pm$1.30 & 79.17$\pm$1.77 & 86.67$\pm$1.32 & 84.17$\pm$1.29 & 70.83$\pm$2.38 & \second{87.83$\pm$1.02} & \best{88.33$\pm$1.18} & 87.50$\pm$1.25 \\
Ovarian  & 66.10$\pm$1.53 & 60.98$\pm$2.58 & 67.38$\pm$1.53 & 69.30$\pm$1.61 & 57.57$\pm$2.91 & 74.41$\pm$1.51 & \second{75.48$\pm$1.56} & \best{76.33$\pm$1.30} \\
Fetus    & 83.60$\pm$0.43 & 86.25$\pm$0.27 & 91.96$\pm$0.11 & 92.05$\pm$0.18 & 86.47$\pm$0.23 & 93.50$\pm$0.11 & \second{93.74$\pm$0.13} & \best{93.83$\pm$0.10} \\
\midrule
Avg      & 78.50 & 77.44 & 81.14 & 81.02 & 75.46 & 84.16 & \best{85.51} & \second{84.91} \\
\bottomrule
\label{tab4}
\end{tabularx}
\end{table*}

The distribution of UniUS-Bench in TinyUSFM feature space (Fig.~\ref{fig5b}) reveals strong generalization capability from pretraining to downstream tasks. Despite training solely on the 3M-US subset, TinyUSFM maintains clear feature separability across diverse downstream datasets, even for organs previously unseen during pretraining. Notably, related anatomical structures exhibit spatial proximity in the embedding space (e.g., breast classification and segmentation tasks cluster closely), while visually similar organs remain well separated (e.g., ovarian and uterine tasks show distinct boundaries). This demonstrates that TinyUSFM learns transferable ultrasound-specific representations with strong discriminability, enabling effective generalization across diverse datasets, imaging protocols, and clinical applications. These results validate the effectiveness of our novel knowledge distillation framework for lightweight ultrasound foundation model construction and highlight its potential to serve as a universal representation backbone for a wide range of downstream tasks.

\subsubsection{Deployability of TinyUSFM}

To evaluate the practical deployability of TinyUSFM, we analyze its computational efficiency, memory consumption, and inference speed. Benefiting from its compact architecture and efficient distillation framework, TinyUSFM contains only 5.5M parameters and requires 2.16 GFLOPs, corresponding to just 6.36\% and 6.40\% of USFM’s parameters and computation, respectively. When deployed on a single GPU, TinyUSFM demonstrates outstanding efficiency. On an RTX 2080 Ti, using a batch size of 16, the average inference time per image is below 1 ms, and the total GPU memory usage remains under 1 GB. This indicates that the model can run smoothly even on devices with very limited computational resources, without the need for high-end accelerators or large memory. Despite this extreme compactness, TinyUSFM maintains nearly identical performance to its large-scale teacher USFM.

\subsection{Downstream tasks adaptation}
We conduct systematic evaluations of TinyUSFM on UniUS-Bench for two downstream tasks: classification and segmentation.

\subsubsection{Diagnostic Classification across Organs}

TinyUSFM’s features are fed into a linear layer to predict target labels. This simple design ensures efficient downstream adaptation while fully leveraging TinyUSFM's representations. 

Table~\ref{tab4} presents comprehensive classification results across the 8 datasets in UniUS-Bench, comparing TinyUSFM against three categories of baseline methods. 

TinyUSFM significantly outperforms conventional architectures including ResNet50 (78.50\%) and ViT-B (77.44\%), while using minimal parameters and computational resources, demonstrating the effectiveness of foundation model pretraining for ultrasound representation learning. Among lightweight architectures, TinyUSFM demonstrates competitive performance against EfficientNet (81.14\%) and MobileViT (81.02\%). In addition, it improves the vanilla ViT-Tiny's performance from 75.46\% to 84.91\%, representing a 9.45\% improvement. Notably, TinyUSFM achieves remarkable breakthroughs on challenging tasks compared to ViT-Tiny, such as ovarian tumor classification (76.33\% vs 57.57\%), breast tumor diagnosis (87.50\% vs 70.83\%), and liver lesion detection (77.70\% vs 68.92\%). While these lightweight models are specifically designed for computational efficiency, TinyUSFM leverages ultrasound domain knowledge through our distillation framework, highlighting the importance of domain-specific adaptation beyond generic efficiency-oriented designs.

Most importantly, TinyUSFM approaches the performance of its teacher model USFM (84.91\% vs 85.51\%), with only a 0.60\% performance gap while achieving 15.7× parameter reduction, 15.6× computational efficiency improvement, and about 5.6× lower inference latency. It also consistently outperforms another ultrasound foundation model, URFM (84.91\% vs 84.16\%), despite using a much smaller compute budget. The knowledge distillation proves particularly effective on complex diagnostic tasks, even surpassing the teacher model performance on fetal plane classification (93.83\% vs 93.74\%) and ovarian tumor subtype classification (76.33\% vs 75.48\%). This near-optimal knowledge transfer confirms the effectiveness of our proposed distillation framework in compressing ultrasound foundation models without sacrificing task performance.

\begin{table*}[htbp]
\centering
\small
\caption{Segmentation performance comparison on UniUS-Bench. Results are reported as Dice score (\%). Best results are in \uline{\textbf{bold-underlined}}, second best are in \textbf{bold}. Latency is reported as per-image latency at batch size = 64.}
\begin{tabularx}{\textwidth}{l *{9}{>{\centering\arraybackslash}X}}
\toprule
Type & \multicolumn{2}{c}{Conventional Model} & \multicolumn{3}{c}{Lightweight Model} & \multicolumn{3}{c}{Foundation Model} \\
\cmidrule(lr){2-3} \cmidrule(lr){4-6} \cmidrule(lr){7-9}
Model & VM-UNet & \mbox{RWKV-UNet} & SegFormer & SeaFormer & ViT-T & URFM & USFM & TinyUSFM \\
\midrule
\#Params & 27.4M & 24.6M & 13.7M & 13.9M & 6.2M & 101.2M & 101.2M & 6.2M \\
GFLOPs   & 8.15 & 22.08 & 5.07 & 2.53 & 3.45 & 53.07 & 53.07 & 3.45 \\
Latency & 3.21ms & 3.18ms & 1.45ms & 0.89ms & 0.46ms & 3.28ms & 3.28ms & 0.46ms  \\
\midrule
Muscle   & 89.19$\pm$9.01 & 90.77$\pm$8.57 & 90.24$\pm$8.70 & 90.49$\pm$8.81 & 88.41$\pm$9.60 & 91.04$\pm$8.88 & \best{92.04$\pm$7.29} & \second{91.92$\pm$6.91} \\
Kidney   & 90.30$\pm$6.09 & 92.88$\pm$4.74 & 89.98$\pm$6.91 & 91.98$\pm$4.88 & 87.76$\pm$8.91 & 93.03$\pm$5.65 & \best{93.82$\pm$4.19} & \second{93.46$\pm$3.97} \\
Stomach  & 74.93$\pm$25.02 & 72.90$\pm$26.89 & 76.00$\pm$25.35 & 73.67$\pm$27.27 & 61.43$\pm$29.01 & 78.69$\pm$22.37 & \best{81.87$\pm$19.22} & \second{80.30$\pm$21.58} \\
Thyroid  & 82.20$\pm$14.31 & \second{82.99$\pm$13.54} & 80.34$\pm$19.35 & 80.06$\pm$19.22 & 70.52$\pm$21.01 & 82.30$\pm$17.87 & \best{83.82$\pm$14.25} & 82.77$\pm$16.37 \\
Ovarian  & 82.41$\pm$21.02 & 83.57$\pm$20.65 & 79.17$\pm$24.63 & 81.03$\pm$21.12 & 76.91$\pm$21.85 & \second{83.58$\pm$19.99} & \best{84.56$\pm$19.98} & 82.65$\pm$21.91 \\
Breast   & 86.56$\pm$13.28 & 87.63$\pm$12.84 & 86.17$\pm$12.25 & 86.94$\pm$13.80 & 73.71$\pm$23.26 & 86.64$\pm$11.98 & \best{90.10$\pm$9.97} & \second{88.88$\pm$11.05} \\
Neck     & 79.07$\pm$14.38 & 78.85$\pm$14.19 & 73.65$\pm$14.90 & 73.78$\pm$14.90 & 72.22$\pm$17.99 & 77.63$\pm$16.29 & \second{79.25$\pm$15.28} & \best{80.00$\pm$14.31} \\
Lung     & 68.83$\pm$33.11 & \second{71.02$\pm$31.23} & 68.00$\pm$31.75 & 66.77$\pm$32.34 & 64.83$\pm$33.33 & 67.11$\pm$32.10 & \best{71.35$\pm$30.77} & 69.57$\pm$30.91 \\
Pelvis   & 95.82$\pm$3.25 & 96.37$\pm$2.94 & 96.13$\pm$3.01 & 95.85$\pm$3.14 & 96.85$\pm$2.91 & 96.85$\pm$2.91 & \second{97.36$\pm$2.70} & \best{97.44$\pm$2.63} \\
Cardiac  & 90.24$\pm$4.98 & 90.28$\pm$5.21 & 90.17$\pm$5.10 & 90.13$\pm$5.12 & 88.77$\pm$6.23 & \second{90.78$\pm$4.75} & 90.46$\pm$4.98 & \best{90.84$\pm$4.80} \\
\midrule
Avg      & 83.96 & 84.73 & 82.99 & 83.07 & 78.06 & 84.77 & \best{86.46} & \second{85.78} \\
\bottomrule
\label{tab5}
\end{tabularx}
\end{table*}

Additionally, the performance gains are consistent across diverse ultrasound imaging scenarios. TinyUSFM shows particularly strong performance on challenging multi-class tasks including fetal plane (93.83\%), which distinguishes six standard fetal views, and breast tumor (87.50\%), which differentiates normal tissue, benign masses, and malignant lesions. The model achieves its most significant improvement on ovarian tumor subtype classification (76.33\%), an eight-class task involving various histopathological types, which is notoriously difficult due to subtle intra-class variations, highly similar texture patterns among subtypes, and substantial class imbalance. For binary classification tasks, TinyUSFM maintains robust performance across different anatomical regions, including carotid artery (84.32\%) and skeletal muscle (91.93\%). The consistent performance across organs and task complexity validates the generalizability of our learned representations and the effectiveness of our feature-gradient driven coreset selection in capturing diverse ultrasound imaging patterns.

\subsubsection{Anatomical Segmentation across Structures}

In the segmentation task, TinyUSFM serves as the backbone to extract multi-scale feature maps from intermediate layers. The extracted features are subsequently processed by a lightweight pyramid neck and decoded through an FPN-style head to produce dense prediction maps. Hierarchical features are extracted from layers {3, 5, 7, 11}, and the fused representation is upsampled to match the input resolution. This design maintains a compact architecture while enabling accurate pixel-wise predictions. 

Table~\ref{tab5} summarizes segmentation performance across ten datasets in UniUS-Bench. TinyUSFM consistently outperforms both conventional segmentation models and lightweight segmentation models. 
Compared to conventional models, our approach surpasses VM-UNet (83.96\%) and RWKV-UNet (84.73\%) by margins of 1.82\%, and 1.05\% respectively. Against lightweight architectures, TinyUSFM shows improvements of 2.79\% over SegFormer (82.99\%) and 2.71\% over SeaFormer (83.07\%), while using fewer parameters and comparable computational resources. Most notably, TinyUSFM significantly outperforms the baseline ViT-Tiny backbone by 7.72\% (85.78\% vs 78.06\%), achieving remarkable improvements on challenging segmentation tasks like gastrointestinal stromal tumor segmentation (80.30\% vs 61.43\%), and breast tumor segmentation (88.88\% vs 73.71\%), validating our ultrasound-specific distillation framework. Furthermore, TinyUSFM approaches the performance of its teacher model USFM (85.78\% vs 86.46\%) with only a 0.68\% gap while achieving 16.3× parameter reduction and 15.4× computational efficiency improvement in segmentation. Remarkably, TinyUSFM even surpasses its teacher model on specific challenging segmentation tasks like neck nerve (80.00\% vs 79.25\%) and cardiac chamber (90.84\% vs 90.46\%), demonstrating that the MIM assisted distillation process can enhance performance on anatomically complex structures. It also outperforms URFM on segmentation overall, achieving 85.78\% average Dice versus 84.77\% as reported, because it distills a stronger and more task-aligned representation from USFM. In addition, the MIM-assisted, data-efficient distillation helps preserve fine-grained boundary details and small-structure cues that are critical for ultrasound segmentation.

The performance gains are consistent across diverse anatomical structures and pathological conditions. TinyUSFM excels in challenging tumor segmentation tasks such as breast tumor (88.88\%), and thyroid nodule (82.77\%). For organ segmentation tasks, robust performance is maintained across organs including kidney (93.46\%), muscle (91.92\%), and cardiac (90.84\%). 

\begin{figure*}[htbp]
    \centering
    \includegraphics[width=0.95\linewidth]{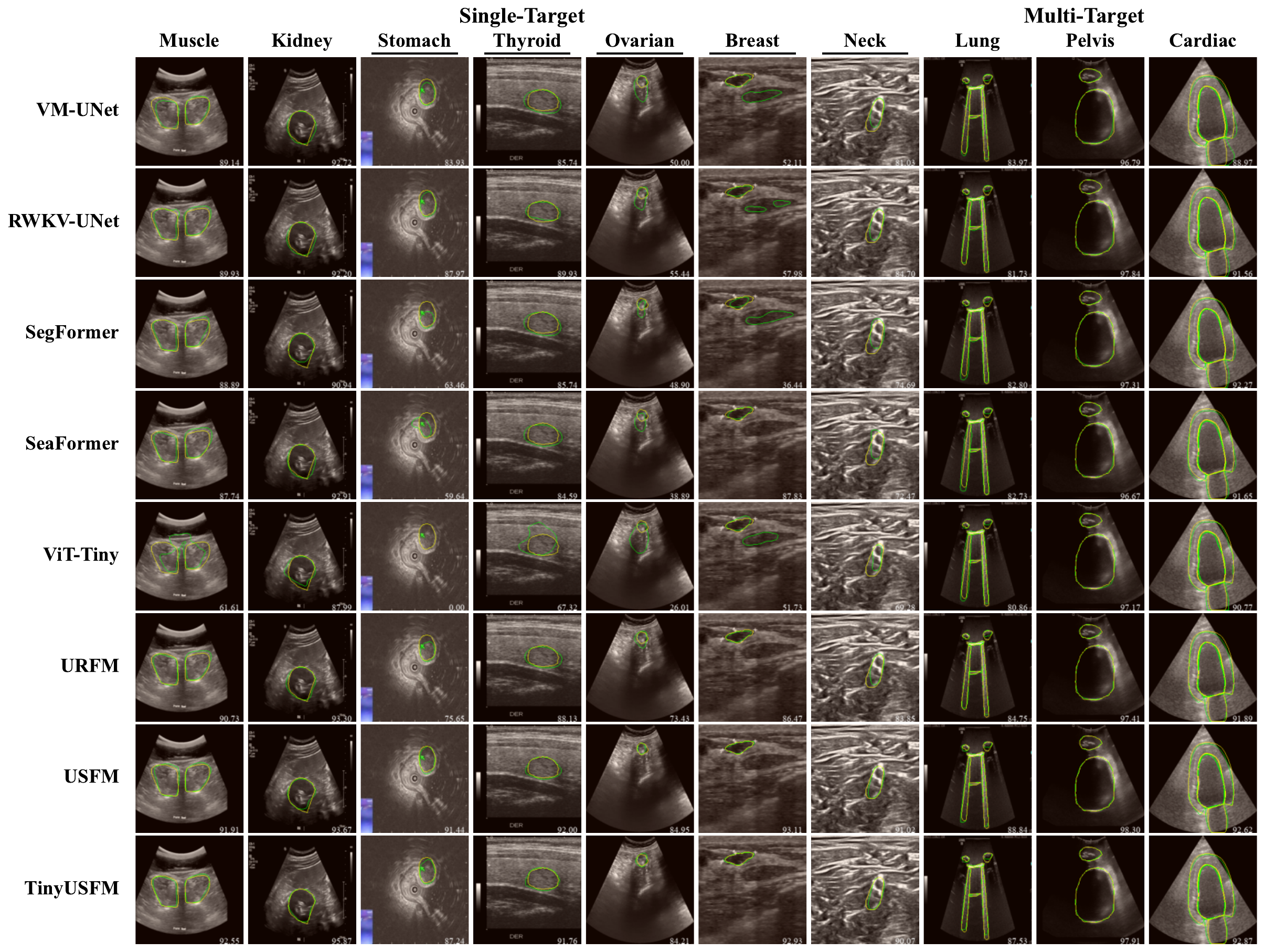}
    \caption{Qualitative comparison of segmentation results across different organs on UniUS-Bench. Depicted in yellow is the ground truth, and in green is the predicted result. Tumor-level datasets are indicated by \uline{underlined titles}.}
    \label{fig6}
\end{figure*}

To provide a visual comparison, Fig.~\ref{fig6} presents segmentation results across ten representative organs from UniUS-Bench. Compared with conventional and lightweight models, TinyUSFM produces more accurate and smoother contours that better align with anatomical boundaries, particularly in structurally complex regions such as the cardiac chambers and neck nerve. The baseline ViT-Tiny exhibits inconsistent delineations and boundary fragmentation, indicating insufficient ultrasound representation learning. By contrast, TinyUSFM maintains high spatial coherence and captures subtle tissue interfaces even under strong speckle noise or low-contrast conditions. The performance advantage is especially evident for tumor-level segmentation (e.g., breast and ovarian lesions), where TinyUSFM yields precise shape preservation and reduced false positives. These qualitative results further confirm that our domain-separated distillation effectively enhances spatial–spectral representation learning and yields anatomically reliable predictions across diverse ultrasound organs and acquisition conditions.

These results demonstrate that our framework effectively transfers ultrasound-specific knowledge to lightweight architectures, surpassing all lightweight models and matching state-of-the-art performance across both classification and segmentation tasks while enabling practical clinical deployment through significantly reduced computational requirements.

\section{Discussion}

The proposed TinyUSFM pioneers a teacher-guided approach for building lightweight foundations in ultrasound imaging. Instead of pursuing architectural compression, it emphasizes data- and representation-centric efficiency through effective knowledge transfer from large teacher model to compact student. This teacher-guided paradigm enables TinyUSFM to retain high representational fidelity with minimal computational cost. The following discussion analyzes how this design narrows the gap with large foundation models, surpasses conventional lightweight networks, and facilitates clinical deployment, followed by limitations and future directions.

\subsection{Tiny model vs large scale foundation model}

A key finding of this study is that carefully distilled lightweight models can approach the performance of large-scale foundation models without inheriting their prohibitive computational demands. Large ultrasound foundation models, such as USFM, rely on hundreds of millions of parameters and dozens of GFLOPs, which makes them unsuitable for deployment in most clinical environments. By contrast, TinyUSFM achieves comparable accuracy and segmentation quality with only 5.5M parameters and 2.16 GFLOPs which is over an order-of-magnitude reduction.

This performance is not a trivial result of model compression, but instead reflects the importance of data curation. Our feature-gradient driven coreset selection demonstrates that a subset of only 200K well-chosen samples can provide richer training signals for lightweight models than the entire 3M-US dataset. This result suggests that lightweight architectures benefit more from quality-focused sampling than from sheer data volume, which often introduces noise and redundancy. The observation that the curated subset even outperforms full-dataset training indicates that lightweight models have different optimization dynamics compared with large-scale networks: whereas large models can tolerate noisy gradients, small models require stable, representative, and information-rich supervision. This insight highlights a paradigm shift for building tiny foundation models, emphasizing data efficiency and representational fidelity rather than brute-force scale.

\subsection{TinyUSFM vs lightweight specific model}

Conventional lightweight models, such as MobileViT and EfficientNet, achieve efficiency by architectural simplification or parameter reduction, but they typically lack adaptability across heterogeneous medical tasks. Their design philosophy emphasizes general-purpose efficiency rather than domain-specific knowledge transfer, leading to limited gains in complex clinical scenarios. TinyUSFM, however, is not just another small network. Its advantage comes from an ultrasound-oriented distillation strategy that explicitly addresses the spectral–spatial complexity of ultrasound signals.

The consistency-driven dynamic distillation selectively emphasizes teacher knowledge that is stable under domain-separated perturbations, preventing the student from inheriting unreliable supervision. Meanwhile, the auxiliary domain-separated MIM reconstruction enables the student to retain mid-level structural and spectral cues that would otherwise be lost during aggressive compression. Ablation studies confirm the complementarity of these components, but more importantly, their integration demonstrates that domain-aware distillation can outperform architecture-centric efficiency. This explains why TinyUSFM not only surpasses vanilla ViT-Tiny by large margins (+9.45\% in classification and +7.72\% in segmentation), but also outperforms efficiency-oriented models designed from scratch. In essence, TinyUSFM illustrates a new design pathway for lightweight foundation models: knowledge transfer tailored to domain characteristics rather than solely architectural reduction.

\subsection{Clinical significance}

The clinical relevance of TinyUSFM lies in bridging the gap between cutting-edge foundation models and real-world deployment constraints. Large-scale models often require specialized hardware, which is inaccessible in most community hospitals and point-of-care settings. By drastically reducing model size and computational demand without compromising diagnostic accuracy, TinyUSFM can be integrated into portable ultrasound scanners, mobile workstations, and edge devices. This not only lowers the barrier to adoption but also democratizes access to advanced AI support in resource-limited regions.
Another dimension of clinical significance is workflow integration. A single lightweight model that supports both diagnostic classification and precise anatomical segmentation reduces the need for maintaining multiple specialized networks. This simplification improves reliability, shortens inference time, and lowers technical complexity for clinicians. Beyond efficiency, TinyUSFM has demonstrated robustness in external validation: winning first place in the MICCAI2025 IUGC Challenge~\cite{ma2025unlabeled},~\cite{bai2026iugc}, where it generalized successfully to unseen datasets. Such independent validation highlights the model’s scalability and reinforces confidence in its clinical readiness. Collectively, these features suggest that TinyUSFM is not merely an academic exercise in model compression, but a practical enabler of foundation-level intelligence in everyday clinical practice.

\subsection{Limitations and future directions}

This work primarily focuses on two-dimensional ultrasound imaging, leaving modalities such as 3D ultrasound, Doppler, or elastography for future exploration. Expanding to these modalities, and to additional tasks like lesion detection or image enhancement, will further test the generalizability of the proposed framework. Nonetheless, the present results already establish a strong foundation for lightweight ultrasound AI in practical settings.

\section{Conclusion}
In this paper, we proposed TinyUSFM, the first lightweight ultrasound foundation model that addresses the challenges of deploying large-scale foundation models in resource-constrained clinical environments. We developed a feature-gradient driven coreset selection strategy and a domain-separated MIM assisted consistency-driven dynamic distillation to enable effective knowledge transfer while preserving ultrasound-specific characteristics. We also established UniUS-Bench, the largest publicly available ultrasound benchmark comprising 8 classification and 10 segmentation datasets across 15 organs. Experimental results on UniUS-Bench demonstrate that TinyUSFM closely matches the performance of the large-scale USFM while requiring significantly fewer parameters and computational resources, maintaining strong generalization across diverse organs, devices, and clinical centers. This work opens new possibilities for deploying efficient ultrasound AI models in point-of-care settings and resource-limited healthcare environments, potentially improving diagnostic accessibility and clinical workflow efficiency.

\section*{References}
\bibliographystyle{ieeetr}
\bibliography{main}

\end{document}